\documentclass[prb,twocolumn,superscriptaddress,showpacs,floatfix]{revtex4}
\usepackage{mathrsfs}
\usepackage{amsmath,amsfonts,amssymb}
\usepackage{epsfig}
\usepackage{graphicx}
\usepackage{wasysym}
\usepackage{bbm}
\usepackage{psfrag}
\usepackage{color}
\usepackage{pstool}
\usepackage{braket}
\usepackage[dvips, bookmarks, colorlinks=true, plainpages = false, citecolor = blue, linkcolor = blue, urlcolor = blue, filecolor = blue]{hyperref}
\newcommand \be{\begin{equation}}
\newcommand \ee{\end{equation}}
\newcommand \bes{\begin{equation*}} 
\newcommand \ees{\end{equation*}}
\newcommand \bea{\begin{eqnarray}}
\newcommand \eea{\end{eqnarray}}
\newcommand \bsea{\begin{subequations}\begin{eqnarray}} 
\newcommand \esea{\end{eqnarray}\end{subequations}}
\newcommand \beas{\begin{eqnarray*}} 
\newcommand \eeas{\end{eqnarray*}}
\newcommand \bfg{\begin{figure}}
\newcommand \efg{\end{figure}}
\newcommand \bfgs{\begin{figure*}} 
\newcommand \efgs{\end{figure*}}
\newcommand \bwt{\begin{widetext}}
\newcommand \ewt{\end{widetext}}

\begin{document}
\title{Topological phases of monolayer and bilayer depleted Lieb lattices}
\author{Arghya Sil}\email{arghyasil36@gmail.com}
\affiliation{Department of Physics, Jadavpur University, 188 Raja Subodh Chandra Mallik Road, Kolkata 700032, India}
\author{Asim Kumar Ghosh}\email{asimkumar96@yahoo.com}
\affiliation{Department of Physics, Jadavpur University, 188 Raja Subodh Chandra Mallik Road, Kolkata 700032, India}

\begin{abstract}
Existence of nontrivial topological phases in a tight 
binding Haldane-like model on the depleted Lieb lattice is reported.
This two-band model is formulated by considering the 
nearest-neighbor, next-nearest-neighbor and
next-next-nearest-neighbor hopping terms 
along with complex phase which breaks the time reversal symmetry
of this semi-metallic system. Topological feature of this model is studied 
along with the presence of sublattice symmetry breaking staggered onsite energy.
Combined effect of these two broken symmetries 
is found crucial for an additional transition between
nontrivial and trivial phases.
System exhibits two types of phase transitions, say, 
between two nontrivial phases and nontrivial to trivial phases.
Nonzero Chern numbers, existence of Hall plateau and
symmetry protected edge states 
confirm the presence of the nontrivial phases.
This two-band system hosts four different types of phases
where two are topological.
Additionally topological properties of 
stacked bilayer of the depleted Lieb lattices
is also studied with similar Haldane-like Hamiltonian.
This four-band system is found to host Chern insulating
phases, with higher values of Chern
numbers supported by in-gap edge states.

 \vskip 1.5 cm
 Corresponding author: Asim Kumar Ghosh 
\end{abstract}
\maketitle

\section{Introduction}
Discovery of topological insulator (TI) has triggered a plethora of
research activity in order to find this bulk insulating phase
within a variety new systems. It begun with the finding of 
Quantum Hall Effect (QHE) by K Von Klitzing\cite{Klitzing}
where strong magnetic field quantized the Hall conductivity of two-dimensional
(2D) electron gas and the topological phase was characterized by 
Thouless-Kohmoto-Nightingale-Nijs (TKNN) invariant\cite{TKNN}.
In another breakthrough, F D M Haldane discovered quantum anomalous Hall
effect (QAHE), where quantization of Hall Conductivity 
was found even in the absence of true magnetic field,
but in the presence of complex hopping terms,
which breaks the time reversal symmetry (TRS)\cite{Haldane}.
In this investigation, Haldane formulated a tight-binding model
on the honeycomb lattice where the  next-nearest-neighbor (NNN)
hopping parameters pick up additional phase
which essentially induces this novel insulating phase.

For 2D systems, this class of insulating phase is characterized by
nonzero values of topological invariant known as Chern number ($C$). 
Systems only with multiple bulk bands could host this particular phase,
where several bands assume definite values of $C$.
In addition the system hosts
 topological edge states in this phase, where the
number of such states is determined by the value of $C$ which is
governed by the bulk-boundary correspondence rule\cite{Hatsugai}.
Order of TIs can be defined in terms of the rule which states
that $n$-th order TI in $d$ spatial dimension gives rise to
 symmetry protected states at ($d-n$)
dimension where $n$ is an integer\cite{BBH1,BBH2,Sil1}.
For the Chern insulating (CI) phase $d=2$, and $n=1$.
Experimental realization of Haldane phase have been made possible by
optical lattices of ultracold atoms\cite{Alba,Shao}.

Investigations are continued afterwards by formulating Haldane-like
tight binding models beyond the honeycomb lattice in order to
find this novel phase in a series of subsequent studies.
As a result, this insulating phase is
observed in other 2D lattices, for examples in 
$\alpha$-$\mathcal T_3$\cite{TGhosh}, dice\cite{Basu}, 
Lieb\cite{Weeks}, kagome\cite{Liu,Guo},
stuffed honeycomb\cite{Sil2}, checkerboard\cite{Sun}, 
star\cite{Chen1,Chen2}, square-octagon\cite{Kargarian,Liu}  
lattices etc.
Side by side, many different classes of TIs have been discovered,
and classified. For example, spin-orbit coupling induced TRS-protected
TI's are characterized by $Z_2$ invariant\cite{Kane}.
Different kind of insulating phases are introduced depending on the
perturbation on the system. For example, in Floquet
topological insulators (FTI),
the system hosts topological phase under photo-irradiation which is otherwise 
trivial\cite{Oka,Sil3}.
In another series of studies, nontrivial topological phase is
found in bosonic magnon excitations which are identified as 
topological magnon insulator (TMI)\cite{Owerre,Sil1}.
This particular phase is observed in ferromagnetic (FM)
Kitaev-Heisenberg models
in the presence of external magnetic field
formulated on honeycomb lattice with spin-anisotropic interaction and 
on CaVO lattice with Dzyaloshinskii-Moriya (DM)
interactions\cite{Moumita1,Moumita2}. The bosonic
magnon dispersion is found topological for both the cases.
FM Heisenberg model on breathing kagome lattice
hosts first as well as second order topological phases\cite{Sil1}.
Existence of topological phases is further observed in
antiferromagnetic (AFM) Heisenberg models formulated on honeycomb and
CaVO lattices \cite{Moumita3,Moumita4}. However, in these models,
bosonic triplet excitation on hexagonal and square plaquettes is
found topological for the respective lattices. 
These are few of the examples in the vast domain of different types 
of TI's. 

In order to frame multi-band systems with single atomic orbital, 
topological models are generally 
formulated on the non-Bravais lattices. 
From the structural point of view, those lattices can be divided into
two different groups. Based on their origin, all those 2D lattices
are derived eventually from either square or triangular lattices which are
Bravais. For examples, Lieb\cite{Weeks},
CaVO\cite{Indrani,Moumita2,Moumita4}, square-octagon\cite{Kargarian,Liu} and
checkerboard\cite{Sun} lattices belong to first group as they are derived from
square lattice either by depleting lattice sites or introducing
nonsymmetric hopping terms in the Hamiltonian. However, their
reciprocal lattice is always a square lattice. 
More precisely, Lieb, CaVO and square-octagon lattices are
nothing but depleted square lattice, where 25$\%$ of sites are
removed from the square lattice in a regular manner for the Lieb
lattice. For both CaVO and square-octagon lattices, 
20$\%$ of the sites are removed. On the other hand,
checkerboard lattice is nothing but (undepleted) square lattice
where tight-binding Hamiltonian
is formulated by invoking nonsymmetric hopping terms.
Finally, symmetry of the resulting Hamiltonian depends
on the nature of depletion and arrangement of hopping paths.

In the same way, honeycomb\cite{Haldane}, stuffed-honeycomb\cite{Sil2},
kagome\cite{Liu,Guo}, breathing kagome\cite{Sil1},
star\cite{Chen1,Chen2}, $\alpha$-$\mathcal T_3$\cite{TGhosh}
and dice\cite{Basu} 
lattices constitute the second group as they 
are derived from the triangular lattice. So their reciprocal lattice
is also triangular. In this group, honeycomb, kagome,
breathing kagome and star 
lattices are depleted triangular lattice while stuffed-honeycomb, 
$\alpha$-$\mathcal T_3$ and
dice lattices are essentially triangular lattice 
with nonsymmetric hopping bonds. Both of them are tripartite also. 
In this study, Haldane-like model is formulated on the
depleted Lieb lattice which is found capable to host
nontrivial topological phases. This depleted Lieb lattice
is constructed from the square lattice by erasing 50$\%$ of its sites. 
Lieb lattice is tripartite square lattice while the depleted one is bipartite. 

On the other hand, a number of investigations have been parallelly carried out
on the Haldane model formulated over 
bilayer systems those yield nontrivial phases, specially on
bilayer graphene systems\cite{Owerre2}.
Topological characterization of such systems facilitates
the emergence of new phases in the expanded parameter
space where the band structure gets 
enriched due to the multiplication of band number by two.
In another study, existence of flat bands with
nonzero Chern number and fractional QHE has been 
reported on twisted bilayer graphene\cite{Sorn}.
In this work, a Haldane model on the bilayer depleted Lieb lattice
has also been formulated where a pair of 
topological phases is found to exist. 
\begin{figure*}[t]
 \centering
 \psfrag{q}{$a\!=\!1$}
 \psfrag{g}{\text {\tiny{$\Gamma$}}}
 \psfrag{k}{\text {\tiny{K}}}
 \psfrag{M}{\text {\tiny{M}}}
\psfrag{kp}{\text {\tiny{K$\prime$}}}
 \psfrag{x}{\text {\scriptsize{$x$}}}
 \psfrag{y}{\text {\scriptsize{$y$}}}
  \psfrag{E}{\text {\tiny {\bf $\delta_3$}}}
 \psfrag{F}{\text {\tiny {\bf $ \delta_4$}}}
 \psfrag{a}{\text{{(a)}}}
 \psfrag{b}{\text{{(b)}}}
 \psfrag{c}{\text{{(c)}}}
 \psfrag{d}{\text{{(d)}}}
 \psfrag{(b)}{\text{\scriptsize{$(b)$}}}
 \includegraphics[width=450pt]{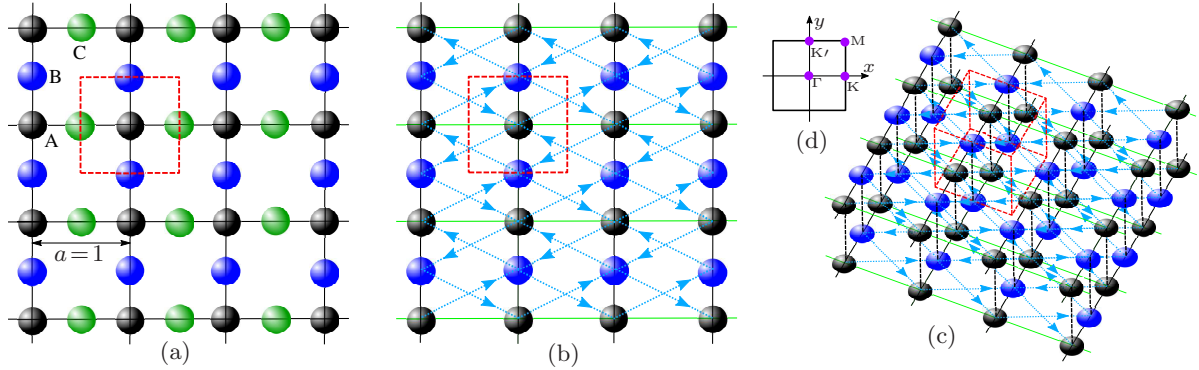} 
\caption{(a) Lieb lattice with three basis points shown by
black (A), blue (B) and green (C) 
spheres. Unit cell is shown by red dashed square.  
(b) Depleted Lieb lattice with two basis points shown by
black (A), blue (B). Strength of NN hopping between A and B
sites along $y$ direction is $t_1$ (black line).
Strength of NNN hopping between A sites along $x$ direction is
$t_2$ (green line), while $t_3$ is that 
along the diagonals between the NNNN lattice sites (blue dotted line).
Sign of the complex phase is assumed positive when 
hopping is along the direction of the arrow. 
(c) Bilayer depleted Lieb lattice, where the hopping strength between
top and bottom layers is $t_4$ (black dashed line).  
(d) The first Brillouin zone (BZ) of the lattice.  High-symmetry points of
square BZ, $\Gamma$ (0, 0), K ($\pi$, 0), K$'$ (0, $\pi$), and
M ($\pi,\,\pi$) have been marked.}
\label{Lattice}
\end{figure*}

Among those several lattice systems exhibiting
CI phase as pointed out above, the
most fascinating one is the Lieb lattice
since it exhibits several exotic topological properties
under various interactions as summarized here.
Tight-binding and Heisenberg
models formulated on Lieb lattice essentially 
yield a three-band systems as it is composed of
three non-equivalent lattice sites.
In tight-binding regime, it shows a flat band\cite{Nita}
and a quadratic band crossing point or a Dirac cone,
which is found topologically protected.
In the presence of spin-orbit coupling (SOC) or incorporation of complex
NNN hopping term, a gap opens up at the Dirac point and the model becomes 
topological\cite{Beugeling,Karnaukhov}. Since the
Lieb lattice possesses inversion symmetry this phase is
characterized by $Z_2$ topological invariant\cite{Weeks}.
FTI phase on Lieb lattice has also been reported in helical photonic
wave-guide\cite{Ren}. TMI phase is 
found in FM Heisenberg model framed on Lieb lattice along with 
DM interaction\cite{Cao}. Stacked 
multi-layer Lieb lattices have been shown to host CI phase with 
high Chern numbers in the presence of SOC\cite{Banerjee}.
Basically Lieb lattice is the 2D
analogue of the 3D lattice found in perovskites. There have been several efforts
to realize materials based on the Lieb lattice. 
This 2D structure has been successfully demonstrated by means of
photonic and cold-atom crystals. In another attempt,
it has been synthesized by the surface state electrons of Cu(111) 
confined by an array of CO molecules positioned
with a scanning tunnelling microscope\cite{Slot}.
More recently, first principle study predicts that two covalent
organic frameworks, sp$^2$C-COF and  sp$^2$N-COF are actually
two material realizations of organic-ligand-based
Lieb lattice\cite{Cui}.

Depleted Lieb lattice
was introduced before by Oliveira {\em et. al.}, 
where a XXZ Heisenberg model was formulated on this lattice. 
The magnetic model comprised of three different nearest-neighbor 
FM exchange interactions\cite{Pires}. However, 
no topological phase was found even in the presence of
DM and external magnetic field. 
In this work, we focus on the depleted
Lieb lattice, and its bilayer structure. But
nontrivial topological phases are obtained by formulating
tight-binding Hamiltonian on those systems. 
This lattice is constructed from the Lieb lattice by removing
few lattice sites in a regular way. A closer look reveals that
it is nothing but the Bravais lattice with rectangular unit cell.
And this particular geometry is formed by the Cu atoms
observed in doped cuprate superconductors with larger
Cu-O-Cu bond length and smaller buckling degree\cite{Kambe}. 
However, a tight-binding model similar to
the Haldane model with nonsymmetric terms is formulated
in such a way that the resulting system can be studied in terms
of non-Bravais depleted Lieb lattice.
In this model next-next-nearest-neighbour (NNNN)
hopping terms are made complex unlike the
original Haldane model where NNN terms are complex instead\cite{Haldane}.
Without the NNNN hopping the system is again
trivial although the presence of staggered site energy
induce a gap in this two-sublattice system. 
Both the complex NNNN hopping and staggered onsite energy 
are the essential ingredients of this model.
The effect of uniform onsite energy would be trivial
for obvious reasons. 

Furthermore, a coupled bilayer version of this
depleted Lieb lattice is introduced where the
coupling among the layers is accomplished by the
inter-layer hopping between same type of nearest-neighbor 
(NN) sublattice sites. It is shown that 
the bilayer system hosts four different
topological phases. 
It is expected that those depleted lattice structures can be
generated using the same techniques employed for
Lieb lattice as discussed before\cite{Slot}. 

The article is organized in the following way.
In section \ref{model}, depleted Lieb lattice is 
described and the tight-binding Hamiltonian is formulated. In section
\ref{Spectra}, spectral properties are described.
This is followed by section \ref{Topology}, where the topological
phases of this monolayer system are presented.
In next section \ref{Bilayer}, topological 
properties of the bilayer version of this lattice are described.
Finally, in section \ref{summary},
discussions and conclusions on this work are summarized.
  
\section{Formulation of Tight-binding Hamiltonian on depleted Lieb lattice}
\label{model}
Geometry of 2D Lieb lattice consisting of three sublattices 
with A (black), B (blue) and C (green) spheres is shown
in Fig. \ref{Lattice}(a).
Unit cell is shown by red dashed square containing the three different sites.
The depleted Lieb lattice is shown in Fig. \ref{Lattice}(b), where
the green sub-lattice points are removed.
This two-sublattice system can be generated alternatively
by depleting blue sub-lattice points instead of green.
The resulting structure is nothing but the rectangular Bravais lattice.
The NN bonds are depicted in black, while the NNN
bonds are green. The NNNN hoppings are
complex as well as nonsymmetric owing to their directionality.
Because of this nonsymmetric bonds the resulting Hamiltonian is 
composed on this non-Bravais lattice. It means adjacent rectangular 
plaquettes along the vertical direction are not identical.
Although the Lieb lattice preserves the four-fold rotational symmetry
($C_4$) of the square lattice, depleted Lieb lattice possesses the
two-fold ($C_2$) rotational symmetry. 
Now, the tight-binding Hamiltonian on this system is formulated as 
\bea
 H\!\!&=&\!\!\! \bigg[t_1\!\sum_{\langle jj'\rangle} c_{j}^{\dagger}\,
c_{j'} +t_2\!\sum_{\langle\langle jj'\rangle\rangle\in A}
  c_{j}^{\dagger}\,c_{j'}\! \nonumber\\
&+& \!\!\!t_3\!\!\!\sum_{\langle\langle\langle jj'\rangle\rangle\rangle}
\!\!\!e^{i\phi_{jj'}}c_{j}^{\dagger}\,c_{j'}\!+ \!H.c \bigg]+
\sum_j \mu_j c_{j}^{\dagger}\,
c_{j}. 
\eea 

The summations $\langle \cdot\rangle$,
$\langle\langle \cdot\rangle\rangle$ and $\langle\langle\langle \cdot\rangle\rangle\rangle$
run over NN, NNN and NNNN pairs, respectively. 
$c_{j}^{\dagger} \,(c_{j})$ is the fermionic creation (annihilation) operator
 for an electron at the $j$-th site. 
 $t_1$, $t_2$ and $t_3$ is the strength of NN, NNN and NNNN hopping,
 those are shown in black, green lines and blue dotted arrow, respectively
 in Fig. \ref{Lattice}(b). 
 $\mu_i$ is the site dependent chemical potential. 
 The direction of the phases $\phi_{jj'}$ is shown in Fig \ref{Lattice} (b).
 The sign of $\phi_{jj'}$ is assumed
 positive, $\phi_{jj'}=\phi$ (negative, $\phi_{jj'}=-\phi$),
 when it is directed along (opposite) the arrowhead.
 Sign of the phases are chosen in this way to ensure the net
 effective magnetic field arising from those phases per square plaquette
 is zero. 
 The onsite energy is taken as $\mu=-\mu_{A}=\mu_{B}$, 
 which breaks sublattice symmetry. Hamiltonian can be expressed as
 \be
H=\sum_{jj'}\psi_j^{\dag} \,g_{jj'}\,\psi_{j'},
 \ee
 where $\psi_j^{\dag}=[c_{Aj}^{\dag}\;c_{Bj}^{\dag}]$,
 $g_{jj'}=g_0I_0+\boldsymbol g\cdot \boldsymbol \sigma$,
 $g_0=t_2\delta_{j,j'\in A}$, $g_x=t_1\delta_{j,j'\in {\rm NN}}+t_3\cos(\phi)\delta_{j,j'\in{\rm NNNN}}$, $g_y=t_3\sin(\phi)\delta_{j,j'\in{\rm NNNN}}$,
 $g_z=-\mu$, $I_0$ is the $2\times 2$ identity matrix
 and $\sigma_\alpha,\,\alpha=x,y,z$ are the Pauli matrices.

\begin{figure}[h]
\psfrag{kx}{ $k_x$}
\psfrag{ky}{ $k_y$}
\psfrag{E}{ $E_\pm(\boldsymbol{ k})/t_1$}
\psfrag{pa1}{$t_1\!=\!1,\,t_2\!=\!3/4,\,t_3\!=\!1/4$}
\psfrag{pa2}{$\phi\!=\!\pi/3,\, \mu\!=\!1$}
\psfrag{ep}{ $E_+$}
\psfrag{em}{ $E_-$}
\psfrag{3}{3}
\psfrag{2}{2}
\psfrag{1}{1}
\psfrag{0}{0}
\psfrag{0.0}{ 0}
\psfrag{-3}{$-$3}
\psfrag{-2}{$-$2}
\psfrag{-1}{$-$1}
\psfrag{p}{ $-\pi$}
\psfrag{p2}{ $-\frac{\pi}{2}$}
\psfrag{pp}{$\pi$}
\psfrag{p3}{$\frac{\pi}{2}$}
  \includegraphics[width=230pt]{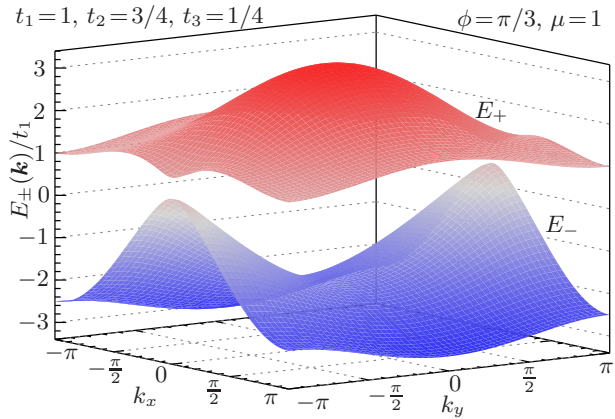}
\caption{Dispersion relations, $E_\pm(\boldsymbol{ k})$, when 
  $t_2/t_1=3/4,\,t_3/t_1=1/4,\,\phi=\pi/3,$
  and $\mu/t_1=1$.}
\label{Dispersion}
\end{figure}

Length of the each side of the square unit cell is set
as the unity ($a=1$). The NN vectors between
A and B sites are $\boldsymbol \delta_1 = \left(0,1/2\right) $,
the NNN vector between the A sites are $\boldsymbol \delta_2 = \left(1,0\right) $ and the NNNN
vectors between the A and B sites along the diagonals
are $\boldsymbol \delta_3 = \left(1,1/2\right)$. 
We set $t_1=1$ throughout the article. 
Hamiltonian is transformed into the reciprocal space
under the Fourier transformation,
 \[c_{j}=\frac{1}{\sqrt N}\sum_{\boldsymbol k\in{\rm BZ}}c_{\boldsymbol k}\,e^{i\boldsymbol k\cdot \boldsymbol R_j},\]
 where $N$ is the number of unit cells,
 $\boldsymbol {k} =(k_x,k_y)$, and $\boldsymbol R_j$
 is the Bravais vector for the site $j$.  
So the Hamiltonian is 
\be H=\sum_{\boldsymbol k}
{\psi_{\boldsymbol k}}^{\dagger}\, h(\boldsymbol {k})\, \psi_{\boldsymbol k}, \ee
where  $\psi_{\boldsymbol {k}}^{\dag} =[c_{A\boldsymbol {k}}^{\dag}\;c_{B\boldsymbol {k}}^{\dag}]$,
is a two-component spinor and
$h(\boldsymbol {k})=h_0I_0+\boldsymbol h\cdot \boldsymbol \sigma$, where, 
\[\left\{\begin{array}{l}
h_0=t_2\cos{k_x},\\ [0.2em]
h_x=2t_1\cos\left(\frac{k_y}{2}\right)+4t_3\cos(\phi)\cos(k_x)\cos\left(\frac{k_y}{2}\right),\\ [0.2em]
h_y=4t_3\sin(\phi)\sin(k_x)\sin\left(\frac{k_y}{2}\right),\\ [0.2em]
h_z=t_2\cos{k_x}-\mu.\end{array}\right. \]
The phase $\phi$ appears here in the off-diagonal terms while
it is found in diagonal terms for the Haldane model\cite{Haldane}.
Conservation of TRS and particle-hole symmetry (PHS) for 
$ h(\boldsymbol {k})$ means that the following relations 
\[\left\{\begin{array}{l}
\mathcal T\, h(\boldsymbol {k})\, \mathcal T^{-1}= h^*(-\boldsymbol {k}),\\ [0.5em]
\mathcal P\, h(\boldsymbol {k})\, \mathcal P^{-1} = -h^*(-\boldsymbol {k}),\end{array}\right. \]
will be satisfied under the appropriate choice of  
TRS and PHS operators, $\mathcal T$, and $\mathcal P$, respectively.
However, no operator is found to satisfy the above relations 
due to all the components,
$h_\alpha$, being nonzero and their typical dependence on
$\boldsymbol k$. As a consequence, chiral symmetry is also broken.
This 2D system belongs to class A, and hence its topological
invariant may assume integral value upon its characterization\cite{Ryu}.
The spectrum of $ h(\boldsymbol {k})$
does not possess the inversion symmetry around the zero energy
because of the absence of PHS. 
Dispersions of this two-band system is given by
\[E_\pm(\boldsymbol {k})=h_0\pm\sqrt{h_x^2+h_y^2+h_z^2}.\]
The dispersions within the BZ are shown in Fig. \ref{Dispersion} for
$t_2/t_1=3/4,\,t_3/t_1=1/4,\,\phi=\pi/3,$ and $\mu/t_1=1$.
It clearly indicates that inversion symmetry for the spectrum of $h(\boldsymbol {k})$
around $E=0$ is lost. A band gap exists for obvious reason.
Inversion symmetry for the spectrum will be restored if $h_0$ 
becomes independent of $\boldsymbol k$ and $h_z=0$. 
The band-structures along the high symmetric pathway
for various parameter values are 
obtained and their properties are described in the next section.

\begin{figure*}[t]
\centering
\psfrag{E}{\scriptsize{$E/t_1$}}
\psfrag{K1}{$k_1$}
\psfrag{K2}{$k_2$}
\psfrag{$3$}{$\pi$}
\psfrag{c0}{\hskip -.1 cm\text{{{$C\!=0\!$}}}}
\psfrag{c1}{\hskip -.1 cm\text{{{$C=1$}}}}
\psfrag{c-1}{\hskip -.18 cm\text{{{$C\!=\!-1$}}}}
\psfrag{G}{\tiny{$\Gamma$}}
\psfrag{K}{\text{\tiny{$K'$}}}
\psfrag{C}{\text{\tiny{$K$}}}
\psfrag{M}{\text{\tiny{$M$}}}
\psfrag{r}{\text{\tiny{$E_3$}}}
\psfrag{s}{\text{\tiny{$E_4$}}}
\includegraphics[width=5.5cm,height=4cm,]{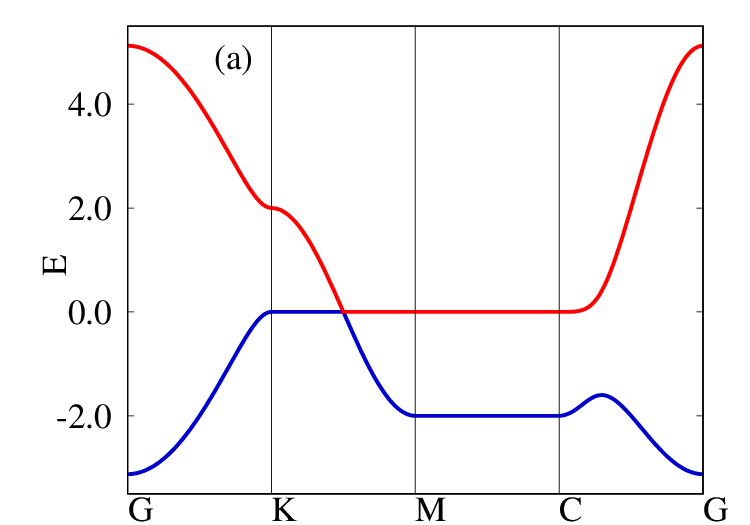} 
\includegraphics[width=5.5cm,height=4cm]{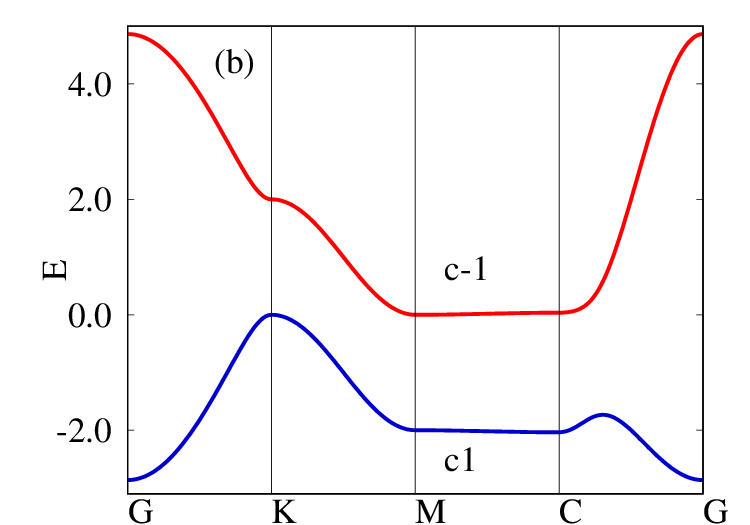}  
\includegraphics[width=5.5cm,height=4cm]{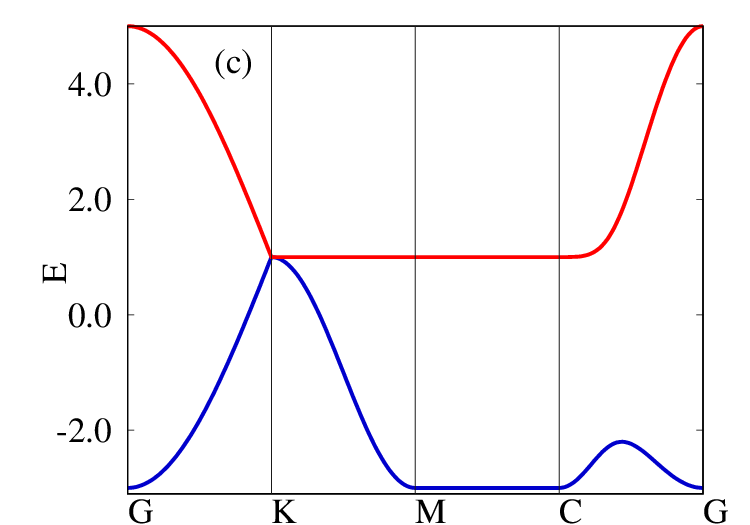}
\includegraphics[width=5.5cm,height=4cm]{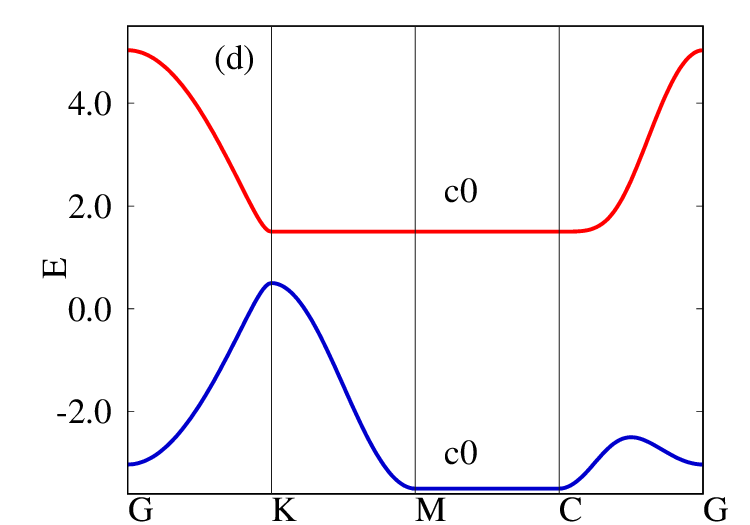} 
\includegraphics[width=5.5cm,height=4cm]{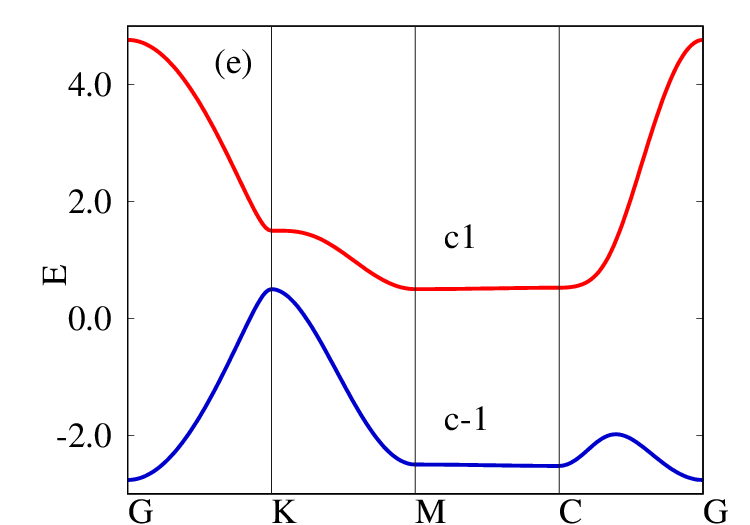}
\includegraphics[width=5.5cm,height=4cm]{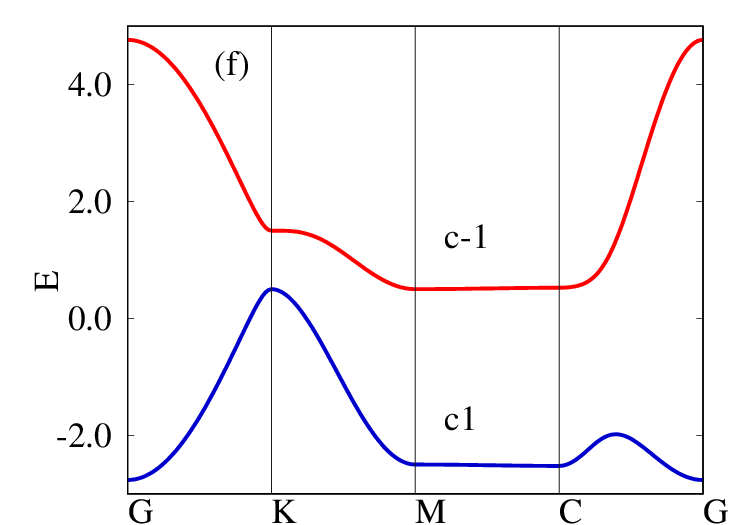} 
\caption{Band structure of depleted Lieb lattice along
the high symmetry pathway $\Gamma$-$K'$-$M$-$K$-$\Gamma$ of the BZ is shown 
with the variation of $\mu$ and $\phi$, when $t_2/t_1=1,t_3/t_1=1/2$. 
(a) $\mu=0$, $\phi=0$, trivial semi-metallic phase with no band gap, 
(b) $\mu=0$, $\phi=\pi/6$, nontrivial semi-metallic phase with pseudo-gap,
(c) $\mu/t_1=1$, $\phi=0$, trivial semi-metallic phase with band-gap point shifting to $K'$ point,   
(d) $\mu/t_1=1.5$, $\phi=0$, trivial insulating phase with true band gap,
(e) $\mu/t_1=0.5$, $\phi=\pi/6$, Chern semimetallic phase with $C_n=(-1,1)$,
(f) $\mu/t_1=0.5$, $\phi=-\pi/6$, Chern semimetallic phase with $C_n=(1,-1)$}
\label{band}
\end{figure*} 
\section{Spectral Properties of depleted Lieb Lattice}
\label{Spectra}
Four different phases of the system are noted, which are shown 
in six band diagrams in Fig \ref{band}(a)-(f). Among them two
phases are topological
as indicated in Fig \ref{band}(b), (e) and (f). In order to explain the
character of them, evolution of spectral properties with the 
variation of complex phase $\phi$ and onsite energy $\mu$ has been demonstrated
in this section. So, values of other parameters are kept fixed, 
namely $t_2/t_1=1$  and $t_3/t_1=1/2$. 
When $\mu$ and $\phi$ are both zero, system is defined only by
NN, NNN and NNNN interactions. In this condition, 
the system remains in one of its trivial state, when the
two bands touch each other at the point $(k_x,\,k_y)=(\pi/2,\,\pi)$,
an intermediate between two high symmetry points $K'$ and $M$, 
as shown in Fig \ref{band} (a). It means the system always exhibit 
semimetallic state when $\mu=0$ and $\phi=0$. 
Now, in the presence of complex NNNN phase, $\phi\ne 0$, but $\mu=0$,
the two bands get separated immediately 
as shown in Fig \ref{band} (b). Again the system remains in semimetallic
phase, since the bands are separated by indirect (pseudo) band-gap.
This phase is termed as pseudo semimetallic phase.
In this case, peak of lower band appears at $K'$
while bottom of upper band remains in $M$-$K$ region. 
The system becomes nontrivial 
as the bands acquire nonzero Chern number. On the other hand, 
when $\mu \ne 0$, but $\phi= 0$, the band-touching point
shifts leftwards with the increase of $\mu$, and at
$\mu/t_1=1$, it moves ultimately to $K'$ point of the BZ (Fig \ref{band}(c)).
It happens due to fact that bottom of upper band
extends with the increase of $\mu$, and it reaches $K'$,
although peak of lower band remains fixed at $K'$. 
This is nothing but another semimetallic phase where band touching
occurs at one of the high symmetry points, $K'$.

With the further increase of $\mu$, or beyond $\mu/t_1=1$, but $\phi= 0$,
the band gap opens up eventually at the $K'$ point and the system 
lies in trivial insulating phase.
The band-structure at the values of 
$\mu/t_1=1.5$ and $\phi=0$ is shown in Fig \ref{band}(d),
which indicates the existence of true band gap. 
Now, when both the values of $\mu$ and 
$\phi$ are nonzero, the system is always in a pseudo-gapped
phase and in most of the cases,
the phase is nontrivial, namely Chern semimetallic (CSM) Phase.
One of the case is demonstrated
in Fig \ref{band} (e). At these values, both the bands acquire
nonzero Chern numbers. The lower (upper) band
acquires $C=-1$ ($C=1$). The sign of the Chern numbers reverses
with the sign reversal of $\phi$ as shown in Fig \ref{band} (f),
leading to another CSM phase. 
The band structures of the last two cases are similar 
since the magnitudes of all the parameters are the same.
However, in every case, it lacks inversion
symmetry about $E=0$, due to obvious reason.
Among the four phases, two are topologically nontrivial, while
the remaining two are trivial. The trivial phases are insulating and 
semimetallic. Two CSM phases emerge here,
although no CI phase is noted in this case.
\section{Topological Properties}
\label{Topology}
\subsection{Chern numbers and Hall conductivity}
As pointed out before, nontrivial topological phases are
characterized by the value of topological invariant, $C$ which
can assume only integral value as it belongs to 2D symmetry class A\cite{Ryu}. 
So, for both CI and CSM phases, each energy band is characterized
by band Chern number, $C_n$, where $n$ is the band index. Again, 
over all the bands, $\sum_nC_n=0$, and 
some of the bands must have nonzero $C_n$ for the system to be nontrivial.
However, for the two-band topological system, each band must
have nonzero value of $C_n$, but opposite in sign. 
In this section,  value of $C_n$ has been determined in order
to characterize the topological phases of this system. 
$C_n$ is defined as the integration of the Berry curvature 
$\Omega_n (\boldsymbol k)$ over the first BZ (1BZ), {\em i.e.}, 
\begin{equation}
 \begin{aligned}
  C_n &=\frac{1}{2\pi}\int_{1BZ}d^2\boldsymbol k \cdot \Omega_n\left(\boldsymbol k \right),
    \label{Cn}
 \end{aligned}
\end{equation}
where $\Omega_n\left(\boldsymbol k \right)=
-i\left(\braket{\partial_{1} u_{n,\boldsymbol k}|\partial_{2} u_{n,\boldsymbol k}}-
    \braket{\partial_{2} u_{n,\boldsymbol k}|\partial_{1} u_{n,\boldsymbol k}}\right)$. 
Here $\ket{u_{n,\boldsymbol k}}$ are the eigenvectors for the
$n$-th band of $h(\boldsymbol k)$ 
and $\partial_{i}=\frac{\partial}{\partial k_i}$. 
$C_n$ is well-defined for a particular band as long as it does not 
touch other neighbouring bands {\em i.e.}, there exists true or pseudo-gap between
the bands. Chern metallic (CM) phase has also been reported before, where
bands overlap each other\cite{Sun}, but in that case Chern number
$C_n$ is not defined and the bands are characterized by Hall conductance.
In our numerical calculation, we use the discretized version of 
Eq. \ref{Cn} introduced by Fukui and others in order to
determine $C_n$\cite{Suzuki}. 

In this model, the presence of TRS breaking NNNN phase, $\phi$  
and sublattice symmetry breaking term $\mu$
gives rise to the non-vanishing Chern numbers. The system is
in nontrivial phase for the entire range $-1<\mu/t_1<1$, and
$-\pi\leq\phi\leq\pi$, except at $\phi=0$.
As soon as $|\mu|/t_1=1$, transition occurs 
from nontrivial to trivial phase, and the Chern numbers of the bands
become zero. For $\phi<0$ ($\phi>0$), the Chern numbers
are $C=(1,-1)$ ($C=(-1,1)$).
So, whenever the value of $\phi$
crosses the $\phi=0$ line for $-1<\mu/t_1<1$, phase transition occurs 
from $C=(-1,1)$ to $C=(1,-1)$ which is accompanied by
closing and reopening of the band gap.

However, Chern numbers for this 2D two-band system
can be obtained analytically using the generalized formula\cite{Sticlet},
\be C_n=\frac{1}{2} \sum_{{\boldsymbol k}\in D_j}{\rm sgn}
(J(\boldsymbol {k})_z) \cdot 
    {\rm sgn}(h(\boldsymbol {k})_z),
\label{Chern-number}
    \ee
where $\boldsymbol J(\boldsymbol {k})=(\partial_{k_x}
    {\boldsymbol h}(\boldsymbol {k})) \times 
    (\partial_{k_y} {\boldsymbol h}(\boldsymbol {k}))$, 
and $D_j$'s are the Dirac points for the Hamiltonian.
Dirac point corresponds to the location in the BZ where doubly degenerate
states exist.  
In this case, by choosing $h(\boldsymbol {k})_z$ as the arbitrary axis, 
Dirac points are obtained by
setting $h(\boldsymbol {k})_x=0$, and $h(\boldsymbol {k})_y=0$. 
At this moment, locations of Dirac points are obtained by the solutions of the
equation, $h(\boldsymbol {k})_z=0$. Thus the
coordinates of the Dirac points are given by 
$D_1=(0,\,\pi)$, and $D_2=(\pi,\,\pi)$ for $\phi \ne 0$,
those are nothing but the
high-symmetry points in BZ, $K'$ and $M$, respectively.
Now  sgn($J_z(\boldsymbol {k})$) and
sgn($h_z(\boldsymbol {k})$) in the parameter space 
can be obtained in a simple way which are
given in the following table when $t_2/t_1=1,$ and $t_3/t_1=1/2$.
\begin{table}[ht]
\centering
\begin{tabular}{|c|c|c|c|}
  \hline
  Conditions & [sgn($h_z$)$\cdot$sgn($J_z$)]$_{_{D_1}}$
  &  [sgn($h_z$)$\cdot$sgn($J_z$)]$_{_{D_2}}$ & $C_n$ \\
  \hline
  $-1\!<\!\mu/t_1\!<\!1$,  &  & & \\
  $0<\phi<\pi$ & $1\cdot 1=1$ & $(-1) \cdot (-1)=1$  & 1\\
   \hline
  $\mu/t_1<-1$,  &  & & \\
   $0<\phi<\pi$ & $1\cdot 1=1$ & $1 \cdot (-1)=-1$  & 0\\
    \hline
  $\mu/t_1>1$,  &  & & \\
    $0<\phi<\pi$ & $(-1)\cdot 1=-1$ & $(-1) \cdot (-1)=1$  & 0\\
 \hline
  $-1\!<\!\mu/t_1\!<\!1$,  &  & & \\
 $-\pi<\phi<0$ & $1\cdot (-1)=-1$ & $(-1) \cdot 1=-1$  & $-1$\\
 \hline
  $\mu/t_1<-1$,  &  & & \\
 $-\pi<\phi<0$ & $1\cdot (-1)=-1$ & $1 \cdot 1=1$  & 0\\
  \hline
  $\mu/t_1>1$,  &  & & \\
    $-\pi<\phi<0$ & $(-1)\cdot (-1)=1$ & $(-1) \cdot 1=-1$  & 0\\
  \hline
\end{tabular}
\caption{Chern number in different parameter regime using
the Eq. \ref{Chern-number}.}
\end{table}

Afterwards it is straightforward to check that in the parameter
regime: $-1<\mu/t_1<1,\, \phi=\pi/6$,
the value of Chern number, $C_n=\frac{1}{2}(1+1)=1$,
and for the parameter regime: $|\mu|/t_1>1,\, \phi=\pi/6$,
the value of Chern number, $C_n=\frac{1}{2}(-1+1)=0$,
and so on.
Evidently, the result matches with the 
numerically calculated values as obtained before. 
Therefore, the system undergoes phase transitions at the points,
$\mu/t_1=\pm 1$, for any nonzero values of $\phi$, where
the transition takes place between nontrivial and trivial phases.
Another transition between two nontrivial phases
occurs around the point $\phi=0$, irrespective of any values of
the remaining parameters. Band gap vanishes at every transition point. 
Obviously coordinates of $D_j$'s depend on the values of parameters. 
\begin{figure}[h]
\centering
\psfrag{k1}{$k_1$}
\psfrag{E}{\hskip -0.2 cm $E/t_1$}
\psfrag{sigma}{$\sigma_H$ ($e^2/h)$ and DOS}
\includegraphics[width=8.0cm,height=5.5cm]{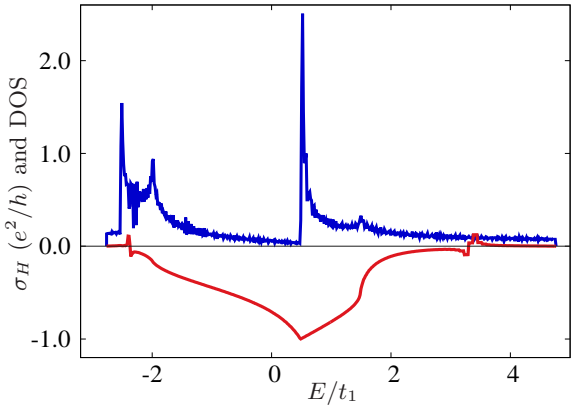} 
\caption{The Hall conductance $\sigma_H$ (red line) and DOS (blue line) 
  with respect to the Fermi energy $E$ for $t_2/t_1=1,\,t_3/t_1=1/2,
  \,\mu/t_1=1/2,\,
  \phi=\pi/6$.
$\sigma_H$ shows a sharp edge instead of a plateau due to
semimetallic phase. It indicates the absence of true band gap.}
\label{Hall}
\end{figure}

The Hall-conductivity ($\sigma_H(E)$) is another important quantity for
determining topological phase 
since it can be verified experimentally. At zero temperature, $\sigma_H(E)$
for this system has been 
estimated numerically by using the Kubo formula\cite{TKNN}, 
\be 
\sigma_H(E)\!=\!\frac{ie^2\hbar}{A_0}\!\!\!\sum_{E_m<E<E_n}\!\!\!\!\!\!\!\!\!
 \frac{\langle m|v_x|n\rangle 
\langle n|v_y|m\rangle\!-\!\langle m|v_y|n\rangle 
\langle n|v_x|m\rangle}{(E_m-E_n)^2},
\ee
where $\ket{l}=\ket{u_{l,\boldsymbol k}}, \,h_{\boldsymbol k}\ket{l}=
E_l\ket{l}$, and $l=m,n$. $A_0$ is the area of the system and 
$E$ is the Fermi energy, while $e$ is the charge of electron. 
The velocity operator, 
$v_\alpha=(1/i\hbar)[\alpha,H]$, where $\alpha=x,y$. 
When $E$ falls in one of the energy gaps, 
the contribution to $\sigma_H$
by the completely filled bands is given by 
\be 
\sigma_H\left(E\right)=\frac{e^2}{h}\sum_{E_n<E}C_n. 
\label{hall-plateau}
\ee
The value of $\sigma_H(E)$ over any  
Hall plateau becomes equal to the sum of all Chern numbers carried by 
the bands having energy lower than it.
The Hall-conductivity shows 
quantized plateau as long as Fermi energy lies in the
band-gap. The height of the plateau is determined by the
Chern number of respective bands according to equation \ref{hall-plateau}.
Variations of $\sigma_H(E)$ and DOS for the
monolayer depleted Lieb lattice with respect to Fermi energy, $E$
are shown in Fig \ref{Hall}. 
As shown in Fig \ref{Hall}, the height of the plateau is
 $\sigma_H=n(e^2/h)$, with $n=-1$ since the Chern number
distribution is $C_n=(-1,1)$. The width of the plateau
tends to zero since the phase is always semimetallic.
$\sigma_H(E)$ shown in Fig \ref{Hall} mostly assumes negative values
due to the topological phase $C_n=(-1,1)$. 
Obviously it will be opposite if it is drawn 
for the other topological phase $C_n=(1,-1)$.

\subsection{Topological Edge States}
\begin{figure}[h]
\centering
\psfrag{y}{\text{\scriptsize{Right edge}}}
\psfrag{z}{\text{\scriptsize{Left edge}}}
\psfrag{kx}{$k_x$}
\psfrag{p}{\hskip -0.01 cm$\pi$}
\psfrag{q}{\hskip -0.27 cm $2\pi$}
\psfrag{wave-square}{\text{ \scriptsize {$| \psi (k_x\!\!=\!\!0)|^2$}}}
\psfrag{Energy}{\hskip -0.4 cm \text{\scriptsize{Energy($k_x$)/$t_1$}}}
\psfrag{site}{\text{\scriptsize{Site}}}
\psfrag{DOS}{\text{\tiny{DOS}}}
\includegraphics[width=8cm,height=7.0cm]{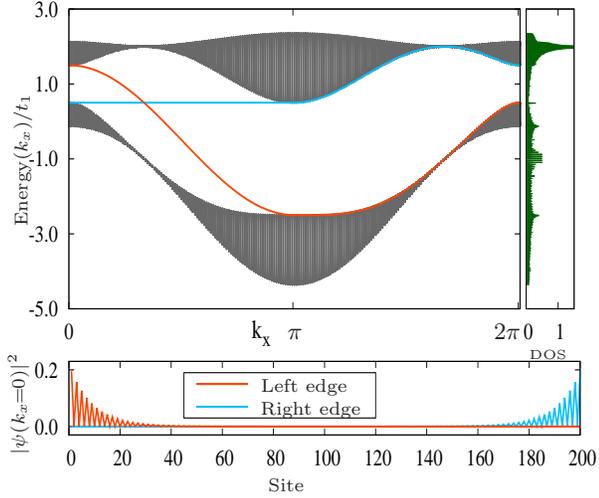} 
\caption{Edge states of depleted Lieb lattice for the 
  parameters $t_2/t_1=1,\,t_3/t_1=1/2, \,\mu/t_1=1/2,\,\phi=2\pi/3$ for $N=200$
  unit cells along $y$ direction.
Chern numbers of the bands are $C_n={-1,1}$. The side-panel 
indicates the density of states. Lower panel shows 
the distribution of probability density of left (orange) and right 
(skyblue) going edge states with respect to site number of 
the strip for $k_x=0$.}
\label{edge}
\end{figure}

`Bulk boundary correspondence' 
rule states that the 
sum of Chern numbers up to the $i$-th band, 
\be
\nu_i = \sum_{j\leq i}C_j,
\label{BBC}
\ee
is equal to the 
number of pair of edge states in the gap\cite{Mook}.
In order to establish this relation,
a 1D ribbon of depleted Lieb lattice is created by removing
the periodic boundary condition (PBC) along $y$ direction so 
that $k_y$ is no longer a good quantum number. We consider $N=200$ sites
along $y$-direction and diagonalize the resulting Hamiltonian
as a function of good quantum number $k_x$, since the
PBC along $x$-direction is present.

In Fig \ref{edge}, the energy eigen-values are plotted 
for $t_2/t_1=1,t_3/t_1=1/2,$ $\mu/t_1=1/2$, $\phi=\pi/6$. Evidently, one pair of 
edge modes exists in the band gap according to 
bulk-boundary rule since the Chern numbers are $(-1,1)$ in ascending
order of energy. Two edge modes connect the two bands but in opposite direction.
The edge states are always localized in the left edge (orange curves) or 
right edge (skyblue curves) as verified from the lower panel of Fig \ref{edge}.
Edge states are chiral in nature since the states with positive (negative)
group velocity are localized in the right (left) edge.
The side panel shows the value of
density of states (DOS) of the system with ribbon
geometry, which clearly indicates the absence of true (direct) band gap
for those particular parameters.
Tight-binding model identical to the original
Haldane model, comprising complex NNN term
and absence of NNNN hopping could not lead to nontrivial topology for
the depleted Lieb lattice. 
\section{Bilayer depleted Lieb Lattice}
\label{Bilayer}
\subsection{Formulation of Hamiltonian}
\begin{figure}[t]
  \centering
  \psfrag{p}{\hskip -0.01 cm$\pi$}
\psfrag{q}{\hskip -0.27 cm $2\pi$}
  \psfrag{E}{\hskip -0.0 cm $E/t_1$}
\psfrag{c0}{\hskip -0.3cm$C\!=0\!$}
\psfrag{c2}{\hskip -0.3cm$C=2$}
\psfrag{c-2}{\hskip -0.3cm$C\!=\!-2$}
\psfrag{G}{\scriptsize{$\Gamma$}}
\psfrag{K}{\text{\scriptsize{$K\prime$}}}
\psfrag{C}{\text{\scriptsize{$K$}}}
\psfrag{M}{\text{\scriptsize{$M$}}}
\psfrag{r}{\text{\scriptsize{$E_3$}}}
\psfrag{s}{\text{\tiny{$E_4$}}}
\psfrag{y}{\text{\scriptsize{Right edge}}}
\psfrag{z}{\text{\scriptsize{Left edge}}}
\psfrag{kx}{\text{\scriptsize{$k_x$}}}
\psfrag{wave-square pi}{\text{ \scriptsize {$| \psi (k_x\!\!=\!\!\pi)|^2$}}}
\psfrag{Energy}{\hskip -0.4 cm\text{\scriptsize{Energy($k_x$)/$t_1$}}}
\psfrag{site}{\text{\scriptsize{Site}}}
\psfrag{DOS}{\text{\tiny{DOS}}}
\includegraphics[width=8.0cm,height=5.0cm,]{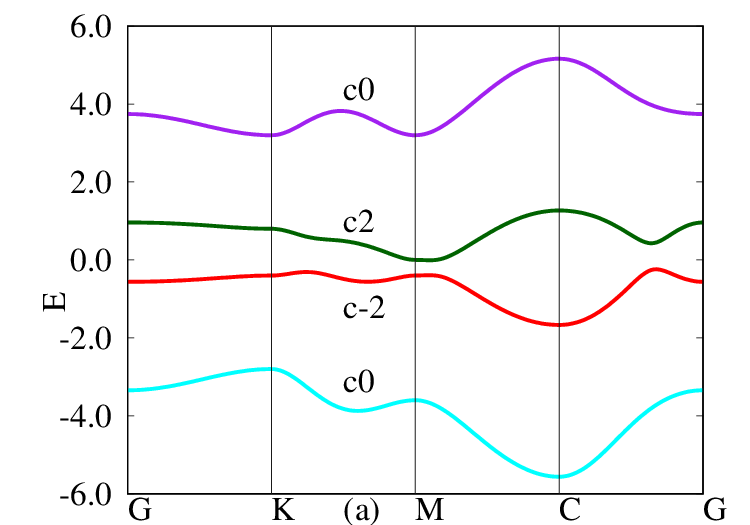} 
\includegraphics[width=8.0cm,height=7.0cm]{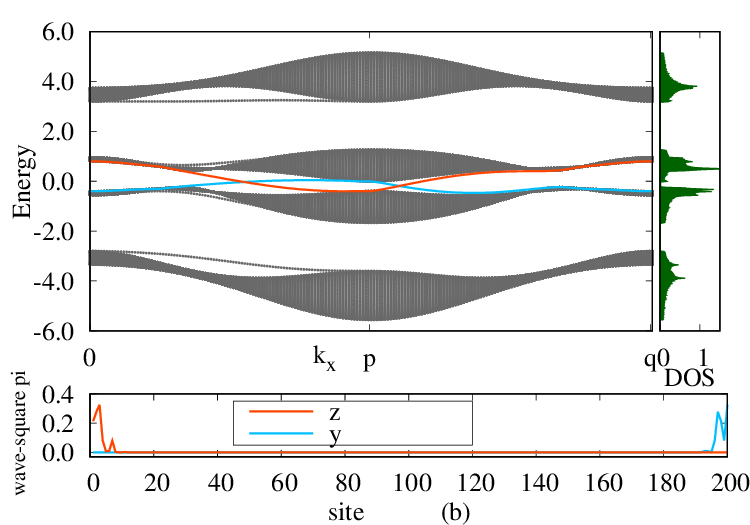}   
\caption{(a) Band structure along the high symmetry points 
of BZ for the bilayer depleted Lieb lattice 
when $t_2/t_1=0.2,\,t_3/t_1=0.5,\,t_4/t_1=1.8,\,
\mu/t_1=1.4,\,\phi^b=2.8,\,\phi^t=1.6$.
(b) Edge states of the bilayer depleted Lieb lattice ribbon for the same
parameters for $N=200$ unit cell along $y$ direction.
Chern numbers of the bands are $C_n=(0,-2,2,0)$. The side-panel 
indicates the DOS of the ribbon. Lower panel shows 
the distribution of probability density of left (orange) and right 
(skyblue) going edge states with respect to site number of 
the strip for $k_x=\pi$.}
\label{band-edge}
\end{figure}
\begin{figure}[t]
  \centering
  \psfrag{p}{\hskip -0.01 cm$\pi$}
\psfrag{q}{\hskip -0.27 cm $2\pi$}
  \psfrag{E}{\hskip -0.0 cm $E/t_1$}
\psfrag{c0}{\hskip -0.3cm$C\!=0\!$}
\psfrag{c1}{\hskip -0.3cm$C=1$}
\psfrag{c-1}{\hskip -0.3cm$C\!=\!-1$}
\psfrag{G}{\scriptsize{$\Gamma$}}
\psfrag{K}{\text{\scriptsize{$K\prime$}}}
\psfrag{C}{\text{\scriptsize{$K$}}}
\psfrag{M}{\text{\scriptsize{$M$}}}
\psfrag{r}{\text{\scriptsize{$E_3$}}}
\psfrag{s}{\text{\tiny{$E_4$}}}
\psfrag{y}{\text{\scriptsize{Right edge}}}
\psfrag{z}{\text{\scriptsize{Left edge}}}
\psfrag{kx}{\text{\scriptsize{$k_x$}}}
\psfrag{wave-square pi}{\text{ \scriptsize {$| \psi (k_x\!\!=\!\!\pi)|^2$}}}
\psfrag{Energy}{\hskip -0.4 cm\text{\scriptsize{Energy($k_x$)/$t_1$}}}
\psfrag{site}{\text{\scriptsize{Site}}}
\psfrag{DOS}{\text{\tiny{DOS}}}
\includegraphics[width=8.0cm,height=5.0cm]{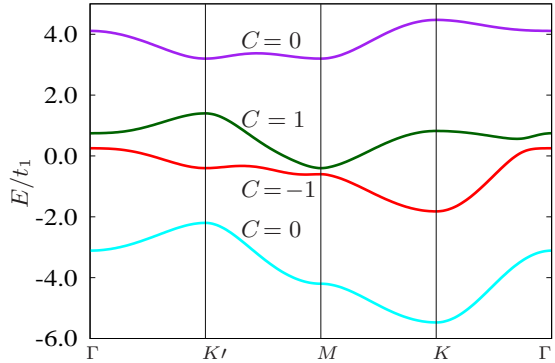}
\caption{Band structure along the high symmetry points 
of BZ for the bilayer depleted Lieb lattice 
when $t_2/t_1=0.5,\,t_3/t_1=0.25,\,t_4/t_1=1.8,\,
\mu/t_1=1.4,\,\phi^b=2.8,\,\phi^t=1.6$.
Chern numbers of the bands from the top are $C_n=(0,1,-1,0)$. }
\label{band-edge-new}
\end{figure}
\begin{figure}[h]
\centering
\psfrag{k1}{$k_1$}
\psfrag{E}{\hskip -0.2 cm $E/t_1$}
\psfrag{sigma}{$\sigma_H$ ($e^2/h)$ and DOS}
\includegraphics[width=8.0cm,height=5.5cm]{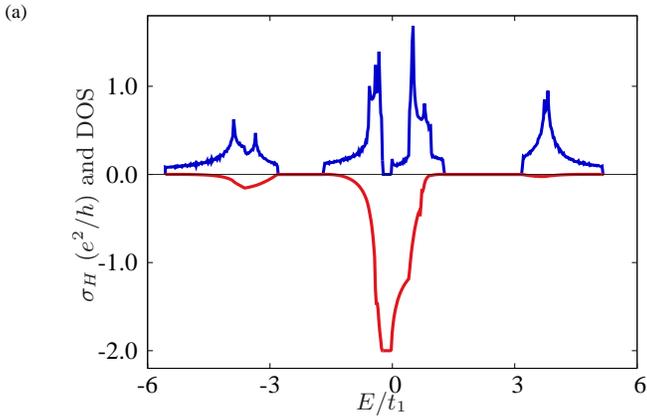} 
\caption{The Hall conductance $\sigma_H$ (red line) and DOS (blue line) 
with respect to the Fermi energy, $E$ for bilayer depleted Lieb lattice
when $t_2/t_1=0.2,\,t_3/t_1=0.5,\,t_4/t_1=1.8,\,\mu/t_1=1.4,\,
\phi^t=2.8,\,\phi^b=1.6$.
$\sigma_H$ shows three plateaus where DOS vanish.}
\label{Hall2}
\end{figure} 
In this section, a system for coupled bilayer of
depleted Lieb lattice is considered, where
two monolayer lattice is stacked identically.
So, the A (B) sublattice of the bottom  
layer lies exactly below the A (B) sublattice of the top layer.
Structure of bilayer depleted Lieb lattice is shown in Fig \ref{Lattice} (c).  
Similar to the monolayer model, the tight binding Hamiltonian is formulated
in the following manner, 
\bea
 H\!\!&=&\!\!\! \sum_{q\in t,b}\bigg[t_1\!\sum_{\langle jj'\rangle} c_{j}^{q\dagger}\,
c_{j'}^{q} +t_2\!\sum_{\langle\langle jj'\rangle\rangle\in A}
  c_{j}^{q\dagger}\,c_{j'}^{q}\! \nonumber\\
&+& \!\!\!t_3\!\!\!\sum_{\langle\langle\langle jj'\rangle\rangle\rangle}
\!\!\!e^{i{\phi_{jj'}^q}}c_{j}^{q\dagger}\,c_{j'}^{q}\!+ \!H.c \bigg]+t_4\bigg[\sum_j c_{j}^{t\dagger} 
c_{j}^{b}+H.c\bigg] \nonumber\\
&+& \sum_{j,q\in t,b} \mu_j c_{j}^{q\dagger}\,c_{j}^{q} . 
\eea 
where $q$ indicates the layer index, which can be `$t$' (top layer) or `$b$' (bottom layer).
 $c_{j}^{q\dagger} \,(c_{j}^{q}),\,q \in t,\,b$, is the fermionic creation (annihilation) operator
 for an electron at the $j$-th site at top ($t$) or bottom ($b$) layer. 
 $t_1$, $t_2$ and $t_3$ is the strength of NN, NNN and NNNN  
hopping respectively and these are same for both the layers. $t_4$ is the strength of inter-layer
hopping between A (and B) sublattices of both the layer.
The direction of the phases $\phi_{jj'}$ is same as in monolayer
case, while ${\phi}^t$
and ${\phi}^b$ can be different. However, choice of 
directions conforms to the fact that total field crossing
the square plaquette is again zero. The onsite energy
is taken as $ \mu_{B}=-\mu_{A}=\mu$, 
for both the layers. None other than NN hopping
between the two layers is taken into account. 

Using PBC along both $x$ and $y$ directions,
the Hamiltonian can be written in the momentum space:
 \be H=\sum_k \psi_{\boldsymbol k}^{\dagger} \,h(\boldsymbol {k}) \,\psi_{\boldsymbol k}, \ee
where $\psi^{\dag}_{\boldsymbol {k}}
= [c^{t \dag}_{A\boldsymbol {k}}\,c^{t \dag}_{B\boldsymbol {k}}\,
c^{b \dag}_{A\boldsymbol {k}}\,c^{b \dag}_{B\boldsymbol {k}}]$,  
is a four-component spinor and $h(\boldsymbol {k})$ is a $4\times 4$ matrix.
The upper diagonal components of $h(\textbf{k})$ are given by 
\begin{equation}
 \begin{aligned}
 h_{11}&= -\mu+2t_2\cos\left(k_x\right) =h_{33}, \\
 h_{12}&= 2t_1\cos\left(\frac{k_y}{2}\right)+4t_3\cos\left({\phi}^b\right)\cos\left(k_x\right)\cos\left(\frac{k_y}{2}\right) \\
 &-4it_3\sin\left({\phi}^b\right)\sin\left(k_x\right)\sin\left(\frac{k_y}{2}\right), \\
 h_{13}&= t_4 =h_{24}, \\
 h_{22}&= \mu =h_{44}, \\
 h_{34}&= 2t_1\cos\left(\frac{k_y}{2}\right)+4t_3\cos\left({\phi}^t\right)\cos\left(k_x\right)\cos\left(\frac{k_y}{2}\right) \\
 &-4it_3\sin\left({\phi}^t\right)\sin\left(k_x\right)\sin\left(\frac{k_y}{2}\right). 
 \end{aligned}
\end{equation}
The other components are zero and the lower diagonal components are complex 
conjugate of the upper diagonal components. Diagonalizing this matrix numerically,
we obtain the band spectrum. In the large parameter-space, there is very few region
where the system is in a insulating or 
semi-metallic phase. In most of the region,
the system is in metallic phase. One of the nontrivial
insulating phase is shown in Fig \ref{band-edge} (a),
in which the middle bands carry nonzero Chern numbers.
Similarly, one nontrivial semimetallic phase has been
identified as shown in Fig \ref{band-edge-new}.
No inversion symmetry about $E=0$ is found in those 
band structures which means PHS is absent in this bilayer
model like the monolayer one.
\subsection{Topological Properties}
Following equation \ref{Cn} and the numerical method developed
by Fukui and others,
we calculate the Chern numbers of the band. Nonzero Chern numbers
are obtained for the
values $t_2/t_1=0.2,\,t_3/t_1=0.5,\,t_4/t_1=1.8,\,
\mu/t_1=1.4,\,\phi^b=2.8,\,\phi^t=1.6$. 
In ascending order of energy, the Chern numbers are given by
$C_n=(0,-2,2,0)$. 
 Hence, the system is in the CI phase for this particular
 set of values.
 Another CI phase with $C_n=(0,2,-2,0)$ is obtained by
 reversing the signs of $\phi^b$ and $\phi^t$, simultaneously. 
 It is interesting to note that, higher Chern number is obtained 
by coupling two identical layers of depleted
Lieb lattice. If we vary the parameters,
the middle two bands overlap and the Chern number becomes undefined.
In addition, one pair of CSM phases with  $C_n=(0,\mp 1,\pm 1,0)$ have been
identified when  $t_2/t_1=0.5,\,t_3/t_1=0.25,\,t_4/t_1=1.8,\,
\mu/t_1=1.4,\,\phi^b=\pm 2.8,\,\phi^t=\pm 1.6$. 
So unlike the monolayer system, both CI and CSM phases
are found in the bilayer system. 

In order to check the `bulk boundary correspondence' 
rule, two isolated edges are created by means of bilayer depleted Lieb
lattice ribbon for the system. The effective Hamiltonian is obtained by 
removing the PBC along $y$ direction. In order to get the edge states,
system of $N=200$ sites along $y$ direction
is considered and the resulting Hamiltonian
is diagonalized where energies obtained as a
function of good quantum number $k_x$.
In Fig \ref{band-edge}(b), the energy eigen-values are plotted 
for the particular set of values.
In this occasion, two pairs of in-gap edge modes
are found to exist since the middle two bands carry the
Chern numbers, either $C_n=\pm 2$, or  $C_n=\mp 2$,
depending on the sign of $\phi$'s. 
The chiral and localized nature of edge states are also verified from the plot.
The side-panel shows the DOS of the ribbon which reveals the existence of
true band gap.
Variations of $\sigma_H(E)$ and DOS for the
bilayer depleted Lieb lattice with respect to Fermi energy, $E$
are shown in Fig \ref{Hall2}, 
when $t_2/t_1=0.2,\,t_3/t_1=0.5,\,t_4/t_1=1.8,\,
\mu/t_1=1.4,\, \phi^t=2.8,\,\phi^b=1.6$.
Hall conductivity shows three
plateaus where DOS vanishes. Two plateaus are found when $\sigma_H(E)=0$,
and one at $\sigma_H(E)=-2$. In contrast no plateau is found for the
monolayer system, since true band gap is absent there.
The width of the plateau depends on the width of the band gap.
As shown in Fig \ref{Hall2}, the height of the plateau is
 $\sigma_H=n(e^2/h)$, with $n=-2$ since the Chern number
distribution is $C_n=(-2,2)$. 
\section{Summary and Discussions}
 \label{summary}
 Search of topological phases in various materials became an
 intriguing part of investigation in condensed matter physics.
 Theoretical discovery of QAHE leads to an enormous
 possibility of finding new topological phases
 in a variety of systems in the absence of true magnetic field\cite{Haldane}. 
 Investigations in this field began by formulating theoretical models
 on specific lattice structures. In the 2D systems, nontrivial topological
phases are found to exist on the models formulated on
non-Bravais lattices. Symmetries of studied models reveal that
those non-Bravais lattices can be derived from either
square or triangular lattices. For examples, 
Lieb\cite{Weeks}, CaVO\cite{Indrani,Moumita2,Moumita4},
square-octagon\cite{Kargarian,Liu} and
checkerboard\cite{Sun} lattics are derived from
square lattice, while  honeycomb\cite{Haldane}, stuffed-honeycomb\cite{Sil2},
kagome\cite{Liu,Guo}, breathing kagome\cite{Sil1},
star\cite{Chen1,Chen2}, $\alpha$-$\mathcal T_3$\cite{TGhosh}
and dice\cite{Basu} 
are derived from triangular lattice. Investigation in this field
continues thereafter by means of finding topological phase
on remaining non-Bravais lattices.

Previous study on the monolayer depleted Lieb lattice
predicted no topological phase while formulating
FM XXZ Heisenberg model with NN, NNN and NNNN exchange interactions 
even in the presence of DM term and external magnetic field\cite{Pires}. 
In contrast, this study lets out the existence of nontrivial topological phases
in monolayer and bilayer structures of depleted Lieb lattice.
This 2D non-Bravais lattice is generated out of square lattice by removing
its sites systematically in such a way that four-fold rotational symmetry 
of original square lattice reduces to two-fold.
The resulting monolayer lattice is bipartite while the bilayer
lattice is quadripartite.

In the nontrivial regime, each of the models host two topological
phases. In each case, tight-binding models are formulated incorporating
NN, NNN and NNNN hopping terms where NNNN terms are complex.
The phases of the complex terms is chosen in such a way that net
magnetic filed per square plaquette vanishes. Although,
model with this characteristic feature does hold topologically
nontrivial phase, additional
staggered onsite energy term is taken into account in order to
note another transition from nontrivial to trivial phase.
System undergoes two types of phase transitions, namely,
between two nontrivial phases and nontrivial to trivial phases.
Topological phases found in monolayer model are characterized by 
$C_n=(\pm 1,\mp 1)$, while
those for bilayer model are defined by $C_n=(0,\pm 2,\mp 2,0)$,
and $C_n=(0,\mp 1,\pm 1,0)$.
CSM phase appears in monolayer model while both CI and CSM phases 
emerge in bilayer case as they are noted before. 
Also as stated before tight-binding model on depleted Lieb lattice
identical to the original Haldane model could not lead to nontriviality. 
It is now established
that any artificial 2D lattices can be realized 
by means of photonic and cold-atom crystals or by
synthesizing the surface state electrons on Cu(111)/CO
substrate\cite{Slot}. Existence of edge states
may be verified in this route. Detection of topological phase
is also possible by observing the feature of $\sigma_H(E)$,
if real materials is found in future. At the same time, search
of new topological phases in the remaining non-Bravais lattices
will be carried on.  
 
\end{document}